\documentclass[twocolumn,aps,preprintnumbers,amsmath,amssymb,floatfix,prl,showpacs]{revtex4}

\usepackage{dcolumn}
\usepackage{bm}
\usepackage{graphicx}

\begin{document}


\title{
Possible Realization of Directional Optical Waveguides in Photonic 
Crystals with Broken Time-Reversal Symmetry
}

\author{F. D. M. Haldane}
\author{S. Raghu}
\affiliation{Department of Physics, Princeton University,
Princeton NJ 08544-0708}

\date{August 30, 2007}

\begin{abstract}
We show how in principle to construct analogs of quantum Hall edge states in
``photonic crystals'' made with non-reciprocal (Faraday-effect) media.
These form ``one-way waveguides'' that allow electromagnetic energy
to flow in one direction only.
\end{abstract}


\maketitle

In this letter, we describe a novel effect involving an interface 
between two magneto-optic photonic crystals (periodic ``metamaterials'' 
that transmit electromagnetic waves) which can 
theoretically act as a ``one-way waveguide'', \textit{i.e.}, 
a channel along which electromagnetic energy can propagate 
in only a single direction, with no possibility of being back-scattered 
at bends or imperfections.  The unidirectional photonic modes confined to 
such interfaces are the direct analogs of the ``chiral edge-states'' 
of electrons in the quantum Hall effect (QHE) \cite{halperin,wen}.  
The key enabling ingredient is the presence of ``non-reciprocal'' 
(Faraday-effect) media that breaks time-reversal symmetry in the 
metamaterial.  

Just as in the electronic case, every two dimensional 
photonic band is characterized by a topological invariant known as 
the Chern number\cite{berrycurv}, an integer that vanishes identically 
unless time-reversal symmetry is broken.  If the material 
contains a photonic band gap (PBG), the Chern number, summed over all 
bands below the gap, plays a role similar to that of the same quantity 
summed over all \emph{occupied} bands in the electronic case.  In particular, 
if the total Chern number \emph{changes} across an interface separating 
two PBG media, there necessarily will occur states localized to the 
interface having a non-zero net 
current along the interface\cite{halperin,wen}.  In the 
photonic case, such states would comprise our ``one-way waveguide''.  

Such an interface between two PBG media can be realized as a domain wall 
in a 2D periodic photonic metamaterial, across which the direction of the 
Faraday axis reverses.  Unidirectional edge states are 
guaranteed in this system provided that the Faraday effect 
generates photonic bands with non-zero Chern numbers.   Here, 
we construct photonic bands with non-zero Chern invariants in a hexagonal 
array of dielctric rods with a Faraday effect.  We then show that as a 
consequence of topology of the single-particle photon bands in the Brillouin 
zone, the edge states of light occur along domain walls (where the Faraday 
effect vanishes).

It may seem surprising that the physics of the QHE
can have analogs in photonic systems.  The QHE is 
exhibited by incompressible
quantum fluid states of electrons - conserved 
strongly-interacting charged fermions - 
in high magnetic fields, while photons are non-conserved  neutral bosons 
which do not interact in linear media;
furthermore,  photonic bands can be described classically, in terms of
Maxwell's equations.    However, the integer QHE can in
principle occur without
any uniform magnetic flux density (just with broken time-reversal
symmetry)  as has explicitly shown  by one of us
in a graphene-like model of 
non-interacting Bloch electrons\cite{haldane88}; thus Landau-level quantization is not an essential 
requirement for the quantum Hall effect.

We have transcribed the
key features of the electronic model of 
Ref.\cite{haldane88} to the
photonic context.  The 
edge-states are a property
of a one-particle eigenstate problem similar to the Maxwell
normal-mode problem, so are replicated in the photonics
problem.  (The QHE itself has no
photonic analog, as it
follows from the Pauli principle of filling
all one-particle states below the Fermi level.)

The Maxwell normal-mode problem in loss-free linear media
with spatially-periodic local frequency-dependent
constitutive relations is a generalized
self-consistent Hermitian eigenproblem, 
somewhat different from the standard Hermitian
eigenproblem. The non-reciprocal parts of
the local Hermitian  permittivity and permeability tensors
$\bm \epsilon(\bm r, \omega)$ and
$\bm \mu(\bm r,\omega)$ are odd imaginary functions of
frequency, so frequency-dependence is unavoidable.
The generalized eigenproblem  has the structure
\begin{eqnarray}
 \bm U^{\dagger}(\bm k)\bm A\bm U(\bm k) |u_n(\bm k)\rangle =
\omega_n(\bm k) \bm B(\omega_n(\bm k)) |u_n(\bm k)\rangle, 
\label{eigen}
\end{eqnarray}
where $\bm U(\bm k)$ is a unitary operator that defines the
Bloch vector $\bm k$; $\bm A$ and $\bm B(\omega)$ are Hermitian
operators, with the real-eigenvalue
stability condition that the Hermitian operator
$\bm B_0(\omega)$ $\equiv$
$(\partial /\partial \omega)(\omega \bm B(\omega)) $ 
is positive definite (this  assumes that the periodic
medium coupled to the electromagnetic fields has a linear
response described by
harmonic oscillator modes, none of which have natural
frequency $\omega_n(\bm k)$ - a detailed derivation has been 
presented in Ref. \cite{sri}).
The eigenfunctions $\langle \bm r |u_n(\bm k)\rangle$ are the 
spatially-periodic
factors of the Bloch states. 
The 
electronic band-structure problem is a simplification of
(\ref{eigen}),
with
$\bm A$ replaced by the one-electron Hamiltonian, $\bm B$ by the
identity operator $\bm \openone$, and $\omega_n$ by the energy eigenvalue.

In this formulation of Maxwell's equations,
the eigenfunction $\bm u_n(\bm k, \bm r)$ 
$\equiv$ $\langle \bm r | u_n(\bm k) \rangle$
is the 6-component 
vector of complex electromagnetic fields
\begin{equation}
\bm u_n(\bm k,\bm r) =
\left (
\begin{array}{c}
\bm E_n(\bm k, \bm r) \\
\bm H_n(\bm k, \bm r)
\end{array}
\right ) .
\end{equation}
In this basis, $\bm U(\bm k, \bm r)$ = $\exp i \bm k \cdot \bm r$
and $\bm A$ = $-i\bm J^a \nabla _a$ (with $\nabla_a$
$\equiv$ $\partial/\partial r^a$, and repeated indices summed),
where
\begin{equation}
 \bm J^a =  
\left (
\begin{array}{cc}
0 & i\bm L^a \\
-i \bm L^a  & 0 
\end{array}
\right )
, \quad
\bm B =
\left (
\begin{array}{cc}
\bm \epsilon(\bm r,\omega) & 0\\
0  & \bm \mu(\bm r,\omega) 
\end{array}
\right );
\end{equation}
here
$\bm L^a$ are the $3\times 3$ spin-1 matrices,
$(L^b)^{ac}$ = $i\epsilon^{abc}$.
If the physical electromagnetic
fields are given by the real parts of
$\bm u_n(\bm k,\bm r)\exp  i(\bm k\cdot \bm r - \omega_n(\bm k)t)$, the
spatially-periodic time-averaged energy density  and energy
current are the
quadratic forms $(\bm u^*_n,\bm B_0(\omega_n)\bm u_n)$,
$(\bm u^*_n,\bm J^a\bm u_n)$.
For our purposes, the key photonic band-structure
quantity is the \textit{Berry connection}
$\mathcal A^a_n(\bm k)$, a real function of $\bm k$
given by
\begin{equation}
\mathcal A^a_n =
\frac{
 \langle u_n|\bm B_0(\omega_n)|\nabla^a_k u_n\rangle -
 \langle \nabla^a_k u_n|\bm B_0(\omega_n)|u_n\rangle }
{2i \langle u_n|\bm B_0(\omega_n)|u_n\rangle} ,
\label{berryem}
\end{equation}
where $\nabla_k^a$ $\equiv$ $\partial/\partial k_a$ is the
$k$-space derivative.  
We obtained (\ref{berryem}) as a generalization of 
the $\bm B$  =  $\bm \openone$
expression\cite{berrycurv}  
by deriving 
(\ref{eigen}) from
a standard Hermitian eigenproblem where electromagnetic
fields are explicitly coupled
to harmonic-oscillator degrees of freedom of the medium \cite{sri}.

The solution of the normal-mode eigenproblem only determines
$\bm u_n(\bm k,\bm r)$ up to an arbitrary $\bm k$-dependent phase factor;
if the replacement 
$\bm u_n(\bm k,\bm r)$ $\rightarrow$  
$\bm u_n(\bm k,\bm r)\exp i \chi_n(\bm k)$
is made,
$\mathcal A^a_n(\bm k)$ $\rightarrow$
$\mathcal A^a_n(\bm k) + \nabla_k^a\chi_n(\bm k)$.
The Berry connection is a ``gauge-dependent'' analog of
the electromagnetic vector potential; the associated gauge-invariant function
(analogous to the magnetic flux density)
is the $k$-space \textit{Berry curvature} $\mathcal F_n^{ab}(\bm k)$ =
$\nabla^a_k\mathcal A_n^b - \nabla^b_k\mathcal A_n^a $.  The 
\textit{Berry phase} $\exp i \phi_n(\Gamma)$ =
$\exp i \oint \mathcal A^a_n dk_a $ associated\cite{berryphase}
 with adiabatic
evolution  around a closed path $\Gamma$
(here in $k$-space) is the 
gauge-invariant analog
of the Bohm-Aharonov phase factor, and can be expressed in terms of the
integral of $\mathcal F^{ab}_n$ over a surface bounded by $\Gamma$  
\cite{berrycurv}.

The Berry curvature satisfies a $k$-space analog of the Gauss law, except
that ``monopole'' singularities emitting total ``Berry flux'' $\pm 2\pi$
can occur at $k$-space points where there are
``accidental degeneracies'' between bands (this quantization of the monopole
charge ensures that the expression for the Berry phase in terms of
Berry curvature on a surface bounded by $\Gamma$ is independent of how
that surface is chosen\cite{berrycurv}).   The integer Chern invariant associated with
any compact surface (2-manifold)  $\Sigma$ in $k$-space is 
\begin{equation}
C_n^{(1)}(\Sigma) = \frac{1}{2\pi}
\iint_{\Sigma} dk_a\wedge dk_b \, \mathcal F^{ab}_n.
\end{equation}
In the case of a 2D band-structure, $\Sigma$ may be taken to be the
2D Brillouin zone (BZ) itself, and $C_n^{(1)}$ is a property of the 2D 
band\cite{tknn}.
If time-reversal symmetry is unbroken, $\mathcal F^{ab}_n(-\bm k)$ =
$-\mathcal F_n^{ab}(\bm k)$, and  Chern numbers vanish.

We now wish to construct a 2D photonic bandstructure where some bands
have  a non-zero
Chern number.   
The key idea is to start with a bandstructure
that has both time-reversal symmetry \textit{and} inversion symmetry,
which allows the existence of
pairs of ``Dirac points''  in the 2D BZ. These
are isolated points where two bands become degenerate, but split apart with
a linear dispersion (resembling that of the massless Dirac equation)
for nearby Bloch vectors.     Dirac points are generically
allowed because if inversion symmetry is present, $\mathcal F_n^{ab}(-\bm k)$
= $\mathcal F_n^{ab}(\bm k)$; in combination with time-reversal symmetry,
this  means that $\mathcal F_n^{ab}(\bm k)$ = 0.    It is then possible
to chose a phase convention such that the eigenfunctions $\bm u_n(\bm k,\bm r)$
of (\ref{eigen}) are real for all $\bm k$.     The eigenproblem
is then of the  real-symmetric type, where it is possible to
find an ``accidental'' degeneracy between two bands by varying just two
parameters (in this case, the 2D Bloch vector $\bm k$);
in contrast, in the general complex-Hermitian case, \textit{three} 
parameters must be varied to find a degeneracy, which cannot be done
by merely ``fine-tuning'' a 2D $\bm k$.      

Dirac 
points can exist in a 2D bandstructure with both spatial-inversion and 
time-reversal symmetry, but
a gap opens if either  symmetry is broken.
While breaking of inversion symmetry 
leads to non-zero Berry curvature (and hence corrections to the ``semiclassical''
equations 
for the trajectories  of light rays in adiabatically-varying
media\cite{nagaosa}), it does not lead to non-trivial
topology of the bands.     In contrast, when a gap opens at Dirac points
due to time-reversal breaking, \textit{the two bands that split apart
inevitably
acquire non-zero Chern numbers}.

We can now give an  in-principle demonstration 
that ``one-way waveguides'' can be constructed using non-reciprocal
photonic crystals. 
Consider a system with a uniform isotropic permeability tensor  
$\mu_0 \delta^{ab}$, 
and an isotropic but spatially-varying permittivity tensor $\epsilon(\bm r)
\epsilon_0\delta^{ab}$, with
\begin{equation}
\epsilon(\bm r) = \epsilon \left ( 1+\lambda
V_G(\bm r)\right ), \quad
V_G(\bm r) = 
2\sum_{n=1}^3 \cos ( \bm G_n \cdot \bm r) ,
\end{equation}
where  $\bm G_n$, $n = 1,2,3$, are three equal-length reciprocal
vectors in the $xy$ plane,
rotated 120$^{\circ}$ relative to each other.
For small $\lambda$, this problem can be solved analytically
in a ``nearly-free-photon'' approximation.
This system has continuous translational invariance
in the $z$-direction, and we will restrict
attention to wavenumbers $k_z$ = 0, with
Bloch vector $\bm k$ in the $xy$ plane.
The electromagnetic fields then
separate into decoupled ``TE'' and ``TM'' sets,
$\{E_x, E_y, H_z\}$ and $\{H_x,H_y,E_z\}$; 
we specialize to the
TE set.  The six corners of the (first)  BZ are at $\pm \bm K_n$, where
$\bm K_1$ = $(\bm G_2-\bm G_3)/3$,
\textit{etc.}, and $|K|$ = $|G|/\surd 3$; since $\bm K_2-\bm K_1$ = $\bm G_3$, the three wavevectors
$\bm K_i$ are equivalent as Bloch vectors. 
To leading order in  $\lambda$ and the  deviation
$\delta \bm k$ = $\bm k - \bm K_i$ of
the 2D Bloch vector from the the BZ corner, 
the three ``free photon'' TE plane waves with
speed $c_0$ split into a
``Dirac-point'' doublet with $\omega$ = 
$\omega_D \pm v_D|\delta \bm k|$ + $O(|\delta \bm k|^2)$,
where
$\omega_D$ = $c_0|K|(1-\lambda/4 + O(\lambda)^2)$, 
$v_D$ = $c_0/2 + O(\lambda)$,
and a singlet $\omega$ = $\omega_0 + O(|\delta \bm k|^2)$,  $\omega_0$ = 
$c_0|K|(1+\lambda/2 + O(\lambda^2))$.

We now perturb the Dirac points by a
Faraday term (which explicitly breaks time-reversal symmetry), with an axis normal to the $xy$ plane,
added to the permittivity
tensor: $\epsilon^{xy}$ = 
$-\epsilon^{yx}$ = $i\epsilon_0\epsilon\eta(\bm r,\omega)$,
where
\begin{equation}
\eta(\bm r,\omega)
= \eta_0(\omega) +
\eta_1(\omega)V_G(\bm r) ;
\end{equation}
$\eta_0(\omega)$, $\eta_1(\omega)$ are real odd
functions of $\omega$.
We assume that,
for $\omega \approx \omega_D$, 
$|\eta_0(\omega)|, |\eta_1(\omega)|  \ll |\lambda | \ll 1$, 
with negligible frequency-dependence.
The Dirac points now split, with dispersion
$\omega$ = $\omega_D \pm v_D (|\delta \bm k|^2 + \kappa^2)^{1/2}$,
where,
to leading order in $\eta$,
$\kappa$  = $|K|(\frac{3}{2}\eta_1(\omega_D) -3\lambda \eta_0(\omega_D))$.

For  small $\kappa$,
the Berry curvatures of the upper and lower
$k_z = 0$ bands near the split  Dirac points
are 
\begin{equation}
F^{xy}_{\pm}(\delta \bm k) = \pm {\textstyle\frac{1}{2}}
\kappa\left (|\delta \bm k|^2 + \kappa^2 \right )^{-3/2}  .
\end{equation}
There is a total integrated Berry curvature of
$\pm \pi$ near each Dirac point, giving total Chern numbers
$\pm 1$ for the split bands.   By inversion
symmetry, the Berry curvatures
at the two Dirac points have the same sign; if the gap was opened
by broken inversion symmetry, with unbroken
time-reversal invariance, they would have opposite
sign, and the Chern number would vanish.

We now consider an adiabatically spatially-varying Faraday term parameterized
by a $\kappa(\bm r)$ that is positive in some regions and negative
in other regions.   The splitting of the Dirac points vanishes locally on the
line where $\kappa(\bm r)$ = 0.     
It is necessary that, in the perfectly periodic
structure with $\kappa$ = 0, there are  no photonic modes at other
Bloch vectors that are degenerate with the modes at the
Dirac points.   

\begin{figure}
\includegraphics[width=3.0in]{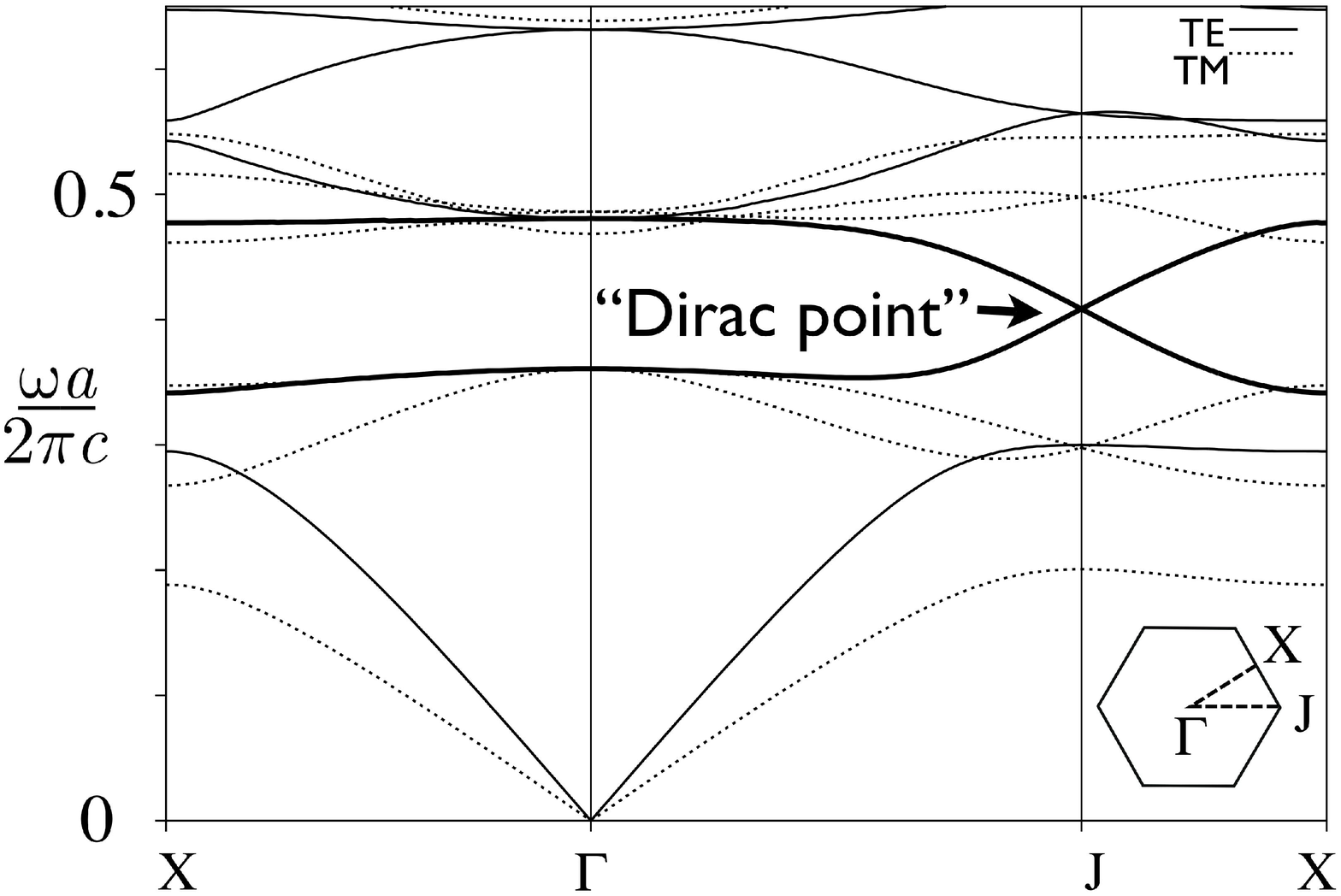}
\caption{\label{fig1}
Photon bands for 
$k_z$ = 0  electromagnetic waves propagating normal
to
the axis of a hexagonal 2D array
of cylindrical
dielectric rods;
$a$ is the lattice constant.
As in Fig.(1a) of Ref.\cite{plihal}, 
the rods fill a fraction $f$ = 0.431 of the volume, with
dielectric constant $\epsilon$ = 14, and are
embedded in an $\epsilon$ = 1 background. 
The lowest five 2D bands are well-separated from higher bands, except near
a pair of ``Dirac points'' at the two distinct
Brillouin zone corners ($J$).
}
\end{figure}

Such frequency-isolation of the Dirac points cannot occur in the weak-coupling
``nearly-free photon'' limit, but can be achieved, at least for
$k_z$ = 0 modes, in hexagonal arrays of
infinitely long dielectric rods parallel to the $z$ axis.
An example can be seen in Fig.(1a) of Ref.\cite{plihal}.
That figure was exhibited to demonstrate a 
frequency gap between the first and second TE
bands, but incidentally also shows
that the second and third TE bands are separated by a
substantial
gap except in the vicinity of the BZ corners, where
they touch at Dirac points.  The corresponding
TM bands were not given in Ref.\cite{plihal}, 
but we found
that the Dirac-point frequency $\omega_D$ is
also inside a  large gap of the TM spectrum 
(see Fig.(\ref{fig1})).  When a Faraday term is 
added, the bands forming the Dirac point in Fig.(\ref{fig1})
split apart, and each now non-degenerate band will have 
associated with it a non-zero Chern number (see Ref. \cite{sri}). 

The Faraday effect incorporated to the hexagonal array of rods explicitly 
breaks time-reversal symmetry on the scale of the unit cell of the metamaterial: the permittivity tensor acquires an imaginary, off-diagonal component having the periodicity of the unit cell, as described above.  A hexagonal array consisting of a material having a large Verdet coefficient, such as a rare-earth garnet with ferromagnetically-ordered domains would give rise to such an effect.  
  
While these $k_z$ = 0 Dirac-point modes are not degenerate with
any other $k_z$ = 0 modes, they \textit{are} degenerate with $k_z \ne 0$ 
modes.   To fully achieve a ``one-way'' edge-mode structure, the light must
also be confined in the $z$-direction, with Dirac points  at a 
frequency that
is non-degenerate with \textit{any} other modes.   
To design such structures, it will be necessary to vary the 
filling factor of the rods along the z-direction, so that light remains 
confined to regions of relatively larger filling factors.  The technical 
challenge would be to vary the filling factors without introducing any modes 
into the bulk TE gaps surrounding the Dirac points.

Let $|u_\sigma(\pm \bm k_D)\rangle$, $\sigma$ = $\pm$,
be the degenerate solutions
of (\ref{eigen}) at a pair of  isolated Dirac points,
normalized so $\langle u_{\sigma}(\pm\bm k_D)|
\bm B_0(\omega_D)|u_{\sigma'}(\pm\bm k_D)\rangle$
= $B_0\delta_{\sigma\sigma'}$.
Now  add a
Faraday perturbation $\delta \bm B(\bm r, \omega)$:
in degenerate perturbation theory, normal modes with
small $\delta \omega$
= $\omega - \omega_D$ have the form 
$\sum_{\sigma,\pm}\psi^{\pm }_{\sigma}(\bm r)\bm U(\pm\bm k_D, \bm r)
\bm u_{\sigma}(\pm \bm k_D, \bm r) $.  For slow spatial variation, there
is negligible mixing between modes at different 
Dirac points, and $\psi^{\pm}_{\sigma}(\bm r)$
is the solution of
\begin{equation}
\sum_{\sigma'}
\left (
-iJ_{\perp}^a \nabla_a
  - \omega_D \delta B(\bm r)
\right )^{\pm}_{\sigma\sigma'}
\psi^{\pm}_{\sigma'}(\bm r) = 
\delta \omega B_0\psi_{\sigma}^{\pm}(\bm r), 
\end{equation}
where $ J^a_{\perp}$ and $\delta B(\bm r)$ are $2 \times 2$ matrices given by 
\begin{eqnarray}
&&\left (J^a_{\perp}\right )^{\pm}_{\sigma\sigma'}
= \langle u_{\sigma}(\pm \bm k_D )|
\bm J^a | u_{\sigma '}(\pm \bm k_D) 
\rangle, \quad a = x,y , \nonumber \\
&&\left (\delta B(\bm r)\right )^{\pm}_{\sigma\sigma'}
= \langle u_{\sigma}(\pm \bm k_D )|
\delta \bm B(\bm r,\omega_D)| u_{\sigma '}(\pm \bm k_D)\rangle .
\end{eqnarray}

For a straight-line interface, this equation has the form
$v_D \hat { K}|\psi\rangle$ = $\delta  \omega |\psi \rangle$,
with $v_D$ $>$ $ 0$, and 
\begin{equation}
\hat K = -i \bm \sigma^x\nabla_x + \delta k_{\parallel}\bm \sigma^y
+ \kappa(x) \bm \sigma^z ,
\label{bound}
\end{equation}
where $\bm \sigma^a$ are Pauli matrices.
Here $k_{Dy}+ \delta k_{\parallel}$ is the conserved Bloch  vector
parallel to the interface; we take
$\kappa(x)$ to be monotonic, with $\kappa(x)$
$\rightarrow$ $\pm \kappa^{\infty}$ as $x$ $\rightarrow $ $\pm \infty$. 

It is instructive to first consider the exactly-solvable case 
$\kappa(x)$ = $\kappa^{\infty}\tanh (x/\xi)$, $\xi > 0$, 
where $\hat K^2$ is essentially the 
integrable P\"oschl-Teller Hamiltonian\cite{LL}. 
The spectrum  of modes bound to the interface is
\begin{subequations}
\begin{eqnarray}
\omega_0(\delta k_{\parallel})
&=& \omega_D +  s_{\kappa} v_D\delta k_{\parallel},
\quad s_{\kappa} \equiv
{\rm sgn}(\kappa^{\infty}),\quad 
\label{zero}
\\
\omega_{n\pm}(\delta k_{\parallel}) &=&
\omega_D \pm v_D
\left ( \delta k_{\parallel}^2 + \kappa_n^2 \right )^{1/2},
\quad n > 0,\quad
\label{bidi}
\end{eqnarray}
\end{subequations}
with $\quad |\kappa_n| < |\kappa^{\infty}|$;
for the integrable model, $\kappa_n^2 $ is
given by $ 2 n |\kappa^{\infty}|/\xi$,
$n$  $<$ $ |\kappa^{\infty}| \xi/2$.
There is always a unidirectional $n$ = 0 mode
with speed $v_D$
and a  direction determined by the sign of $\kappa^{\infty}$; in the
small-$\xi$ (or sharp-wall) limit
$|\kappa^{\infty}| \xi$ $<$ 2, this is the only
interface mode.  

Let
$\phi(\kappa^2)$ be the dimensionless area in the
$x$-$ k_x$ phase-plane enclosed by a closed constant-frequency orbit 
$(k_x)^2 + (\kappa(x))^2$ = $\kappa^2$ $<$ $|\kappa^{\infty}|^2$,
corresponding to a bound state.
For the integrable model, this has the simple form 
$\phi(\kappa^2)$ = $\pi\kappa^2\xi/|\kappa^{\infty}|$; the 
$n > 0$ bidirectional
modes thus satisfy
a constructive-interference condition
\begin{equation}
\phi(\kappa_n^2) = 2\pi n.
\label{quant}
\end{equation}
This contrasts with the usual ``semiclassical'' condition
 $\phi$ = $2\pi(n + \frac{1}{2})$;
the change
is needed for the $n=0$ ``zero mode''
(\ref{zero}) to exist, and can be interpreted as deriving from an extra 
Berry phase 
factor of $-1$ 
because the orbit encloses a Dirac degeneracy
point at $(x,k_x)$ = $(0,0)$.  
For general $\kappa (x)$, the $n$ = 0
eigenfunction is
\begin{equation}
\psi^{0}_{\sigma}(\bm r)
\propto  \varphi_{\sigma}(s_{\kappa})\exp \left (
i\delta k_{\parallel}y - s_{\kappa}\int^x \kappa(x') dx' \right ) ,
\end{equation}
$\bm \sigma^y \bm \varphi(s)$ = $s\bm \varphi (s)$.
For slowly-varying $\kappa(x)$, the condition
(\ref{quant}) 
will determine $\kappa_n^2$ for 
any $n > 0$ interface modes.

Since there are two Dirac points, there are  two such  unidirectional
edge modes at a boundary across which  the Faraday axis reverses.
The crucial feature is that both modes propagate in the 
\textit{same} direction, and cannot disappear, even if the interface becomes
sharp, bent, or disordered.   As in the QHE,
the difference between the number of modes moving in the two
directions along the interface is topologically determined
by the difference of the total Chern number of bands at frequencies
below the bulk photonic band gap in  the regions on either side of the 
interface; in this case $|\Delta C^{(1)}|$ = 2.

For $|\delta \omega|$ $<$ $v_D|\kappa^{\infty}|$
a Faraday interface has no counterpropagating modes into which
elastic backscattering can take place, so the ``one-way waveguide''
that it forms is immune to localization effects, just like electronic
transport in the QHE.    In the QHE, the number of electrons is strictly
conserved; in photonics, the photons only propagate ballistically
if absorption and non-linear effects are absent.  These effects
do allow degradation of the electromagnetic energy current flowing
along the interface, so the analogy with the QHE is not perfect.

Even if a 2D metamaterial with isolated Dirac points can be designed,
the problem of finding a suitable magneto-optic
material to provide the Faraday effect must be addressed.
The effect must be large enough to induce a gap
that overcomes the effect of any inversion-symmetry breaking.
The parameter $|\kappa^{\infty}|$ is the inverse length that controls
the width of the unidirectionally-propagating channel (and the
unidirectional
frequency range); in order to keep the wave confined
to the interface, and prevent leakage,
the Faraday coupling must be strong
enough so that this width is  significantly smaller than
the physical dimensions of the sample of metamaterial.

In summary, we have shown that analogs of quantum Hall effect edge modes 
can in principle occur in two dimensional photonic crystals with broken time-reversal 
symmetry.  The electromagnetic energy in these modes travel in a single 
direction.  Explicit theoretical examples of such modes have been constructed 
in Ref. (\cite{sri}).  
Such quasi-lossless unidirectional channels
are a novel possibility that might one day
be physically
realized
in ``photonic metamaterials'' with
non-reciprocal constituents.

This work was supported in part by the U. S. National
Science Foundation (under MRSEC Grant No. DMR02-13706) at  the Princeton
Center for Complex Materials.
Part of this work was carried out at the Kavli
Institute for Theoretical Physics, UC Santa Barbara,
with support from KITP's NSF Grant No. PHY99-07949.

\end{document}